\documentstyle[12pt]{article}
\newcommand{\la}{\label}

\newcommand{\bfx}{{\bf x}}
\newcommand{\bfy}{{\bf y}}
\newcommand{\bfn}{{\bf n}}

\newcommand{\non}{\nonumber}
\newcommand{\be}{\begin{equation}}
\newcommand{\ee}{\end{equation}}
\newcommand{\ba}{\begin{eqnarray}}
\newcommand{\ea}{\end{eqnarray}}
\newcommand{\bastar}{\begin{eqnarray*}}
\newcommand{\eastar}{\end{eqnarray*}}
\newcommand{\half}{{1 \over 2}}
\begin{document}
\begin{titlepage}
 
\begin{flushright}
hep-th/9705176
\end{flushright}

\vskip 1.0cm 
\begin{center}
{ 
\bf \Large \bf TOROIDAL CONFIGURATIONS \\
                AS STABLE SOLITONS \\ }
\end{center}
 
\vskip 1.0cm
 
\begin{center}
{\bf L. Faddeev$^{* \sharp}$ } {\bf \ and \ } {\bf 
Antti J. Niemi$^{** \sharp}$ } \\
\vskip 1.5cm
{\it $^*$St.Petersburg Branch of Steklov Mathematical
Institute \\
Russian Academy  of Sciences, Fontanka 27 , St.Petersburg, 
Russia$^{\ddagger}$ } \\

\vskip 0.5cm

{\it $^{**}$Department of Theoretical Physics,
Uppsala University \\
P.O. Box 803, S-75108, Uppsala, Sweden$^{\ddagger}$} \\

\vskip 0.5cm
and \\
\vskip 0.5cm

{\it $^{\sharp}$Helsinki Institute of Physics \\
P.O. Box 9, FIN-00014 University of Helsinki, Finland} \\

\end{center}

\vskip 0.7cm
\rm
\noindent
Previously we have proposed that in certain
relativistic quantum field theories knotlike 
configurations may appear as  stable solitons. 
Here we present a detailed investigation of
the simplest knotted soliton, 
the torus-shaped unknot. 

\vfill

\begin{flushleft}
\rule{5.1 in}{.007 in} \\
$^{\ddagger}$  \small permanent address \\ \vskip 0.2cm
$^{*}$ \small Supported by Russian Academy of Sciences
and the Academy of Finland \\ \vskip 0.2cm 
$^{**}$ \small Supported by G{\"o}ran Gustafsson
Foundation for Science and Medicine \\
\hskip 0.3cm and by NFR Grant F-AA/FU 06821-308
\\ \vskip 0.3cm
$^{*}$ \hskip 0.2cm {\small  E-mail: \scriptsize
\bf FADDEEV@PDMI.RAS.RU and FADDEEV@PHCU.HELSINKI.FI } \\
$^{**}$  {\small E-mail: \scriptsize
\bf NIEMI@TEORFYS.UU.SE}  \\
\end{flushleft}

\end{titlepage}
\vfill\eject
\baselineskip 0.65cm

\noindent

\section{Introduction}

Lord Kelvin was the first to suggest that knotlike 
configurations might be of fundamental importance.
In \cite{kelvin} he proposed that atoms, 
which at the time were considered as elementary particles, 
could be viewed as knotted vortex tubes in ether. 
Subsequently he also conjectured \cite{kelvin2}, 
that (thin) vortex filaments that have 
the shape of torus knots \cite{ati}, \cite{kauf} 
should be stable. 

Kelvin's theory of vortex atoms has long ago subsided.
However, at the time it was taken seriously, and it
led to an extensive study and classification of knots. 
In particular the results obtained by  Tait 
\cite{tait} remain a classic contribution to 
mathematical knot theory. Recently Kelvin's 
idea that different elementary particles can be 
identified with topologically distinct knots has 
been advanced in particular by Jehle \cite{jehle}. 

Since its inception, Kelvin's conjecture on 
the stability of torus knots has also attracted 
much interest. However, it seems that 
until very recently \cite{nature}, the issue
has not been even properly addressed. This is 
due to the exceedingly complex nature of dynamical 
models that describe stable knots. 

In order to study the dynamics of 
knotlike configurations in a predictive manner, 
we need a Lagrangian field theory where knots 
can appear as solitons. But in such a theory 
the equations of motion are generically 
highly nonlinear, to the extent that an analytic 
approach is hopeless. Indeed, even
the identification of a Lagrangian 
where stable knotlike  solitons can be present, 
has until recently defied all attempts.   
With the recent, rapid advances in computers 
it appears that by combining topological and
geometrical methods in quantum field theories 
with high performance computing,
a first principles analysis of Kelvin's 
conjecture is finally a reality \cite{nature}. 

In addition of elementary particles (atoms), 
Kelvin was also interested in knotlike 
configurations in a hydrodynamical context. 
Today there are numerous additional 
scenarios, where dynamical knots can
be important: Presently it is commonly 
accepted that fundamental 
interactions are described by string theories
\cite{green}, with different elementary particles 
corresponding to the vibrational excitations of a 
primary string. Even though connections 
between modern string theory and knot theory do
exist, for example the Chern-Simons action is related
both to conformal field theories and knot 
invariants \cite{wit}, the possibility of an intimate 
relationship {\it e.g.} at some nonperturbative
level remains to be investigated. But in a number 
of other physical scenarios 
the potential relevance of knotlike structures 
has already been established \cite{ati}, \cite{kauf}.
For example in early universe cosmology
cosmic strings are supposed to be responsible 
for structure formation. These strings
are expected to decay {\it e.g.} via gravitational
fluctuations, but the possibility that stable knotted
strings might survive should have important 
consequences. Furthermore,
in QCD one may expect that gluonic flux 
tubes that confine quarks in hadrons could become 
tangled, suggesting that in pure Yang-Mills theory 
closed knotted flux tubes appear as physical states. 
Stringy knotted vortices might 
also appear in a variety of condensed matter physics
scenarios. For example in type-II superconductors 
magnetic fields are confined within the 
cores of vortex-like structures. Recent 
experiments with $^3$He-A 
superfluids have also revealed interesting 
vortex structures that can be described by 
theoretical methods which are 
adopted from cosmic string models. 
The study of knotlike 
configurations is also highly important 
for chemical compounds such as 
polymers. Finally, the 
investigation of knots is rapidly becoming 
an important part of molecular 
biology, where entanglement 
of a DNA chain interferes with vital life processes
of replication, transcription and  recombination.

Until now knotlike configurations
have been mainly studied using non-dynamical,
phenomenological models \cite{simon}. 
One first introduces a one-dimensional,
structureless string 
and distributes some charge along it. The configuration 
is then allowed to relax into an equilibrium shape. 
However, such an approach is not dynamical,
there are inevitably several 
knot-specific {\it ad hoc} parameters that need 
to be determined by various means. For
example, the length of each topologically distinct knot
is an independent degree of freedom. The 
details of the charge distribution also
introduces some arbitrariness that can be parametrized. 
As a consequence of such parameters that are 
specific for individual knot configurations,
the ensuing models lack in their 
predictive power. There is a definite
need for a first principles approach where 
different knots emerge dynamically, 
as solitons. One can then predict
their properties in terms of fundamental quantities 
which are in no manner specific to a particular 
knotted structure but characteristics of
the underlying physical environment.

The literature on solitons is enormous,
and there are several extensive reviews \cite{rebbi}. 
Until now the activity has mainly concentrated
on 1+1 dimensions with the notable exceptions of
the 2+1 dimensional vortex and nonlinear $\sigma$-model
solitons, 
and skyrmeons and magnetic monopoles in 3+1 dimensions.
These are all pointlike configurations, that
can not be directly associated with knotted structures.

When a pointlike two dimensional soliton 
is embedded in three dimensions, it becomes a 
line vortex. Since the energy of a vortex
is proportional to its length, for a 
finite energy the length must
be finite. This is possible if the core 
forms a closed, knotted structure.
In 1975 one of us \cite{fadde1} proposed that a 
torus-shaped closed vortex could be constructed in a 
definite dynamical model. The configuration suggested 
in \cite{fadde1} is constructed from a finite length
line vortex, twisted once around its core before joining 
the ends. The twist ensures that the configuration 
is stable against shrinkage.  As a knot it 
corresponds to the {\it unknot} which is 
the simplest possible knotlike configuration \cite{ati},
\cite{kauf}. By 
combining geometrical and topological methods with
high performance computing, we recently verified
\cite{nature} the existence of such a soliton in
the model proposed in \cite{fadde1}. Furthermore,
we also found definite evidence for the existence
of a soliton in the shape of a trefoil, which is
the simplest possible torus knot. This is a strong
indication, that the model proposed 
in \cite{fadde1} realizes Kelvin's conjecture 
on the existence of stable torus knots. 

\vskip 0.4cm 

In this article we shall present a detailed
investigation of the torus-shaped unknot soliton
in the model proposed in \cite{fadde1}. In the 
next section we describe this model in detail,
and in section 3. we
specify it for torus-shaped solitons. 
In section 4. we reformulate
the equations of motion so that a numerical 
construction of the unknot soliton can be performed, and 
in section 5. we present results of a model numerical 
simulation.

\vskip 1.0cm

\section{A Hamiltonian with Localized Solitons}

In \cite{nature} we have
proposed that certain relativistic field
theories may admit solitons with
the topology of a generic torus knot. In particular, 
we argued that such solitons can be described 
in terms of a three component vector 
field $\bfn ( \bfx) $ with unit length $ 
\bfn \cdot \bfn = 1 $. The
action is \cite{fadde1}
\be
S \ = \ \int d^4x \Biggl( \frac{1}{2e^2}
(\partial_\mu {\bfn})^2 \ + \ \frac{1}{4g^2} 
( \bfn \cdot  \partial_\mu \bfn \times 
\partial_\nu \bfn )^2 \Biggr) 
\la{lagrangian}
\ee
where we choose $\bfn$ to be dimensionless 
so that $e$ determines a length scale and
$g$ is a dimensionless coupling constant. 
This is the most general
relativistically invariant action for $\bfn$
that involves second and fourth order
derivative terms so that time derivates appear
only quadratically. It can be viewed as a U(1)
gauged version of the Skyrme model \cite{rebbi}.
However, the dynamical 
content of these two models are quite different.

Since time derivatives in action (\ref{lagrangian})
appear only quadratically, it admits a 
canonical interpretation with static 
Hamiltonian
\be
H \ = \ E_2 + E_4
\ = \  \int d^3x \Biggl( \frac{1}{2e^2}
(\partial_i {\bfn})^2 \ + \ \frac{1}{4g^2} 
( \bfn \cdot  \partial_i \bfn \times 
\partial_j \bfn )^2 \Biggr) 
\la{ham1}
\ee
The first term $E_2$ coincides with the 
nonlinear O(3) $\sigma$-model action, known to
admit stable solitons in two dimensions. But if we 
include the fourth order derivative term $E_4$,
stable finite energy solitons are also possible
in three dimensions. This is suggested  
by the Derrick scaling argument: If we set 
\be 
\vec{\bf x} \ \to \ \rho \vec{\bf x}
\la{scalex}
\ee
we find for our Hamiltonian
\be
H \ = \ E_2 + E_4 \ \stackrel{\rho}{\rightarrow} \
\rho E_2 + \frac{1}{\rho} E_4 
\la{scaling}
\ee
so that stable finite energy
solitons may exist when $E_4$ is 
present. In particular, we conclude that
such solutions obey the virial theorem
\be
E_2 \ = \ E_4
\la{virial}
\ee

A soliton described by $\bfn(\bfx)$ 
is a localized
configuration in $R^3$, at spatial infinity 
$| \bfx | \to \infty$ the
vector field $\bfn(\bfx)$ approaches a constant 
vector $\bfn_0$. Hence we
compactify $R^3$ into a 
three dimensional sphere $S^3$, and 
$\bfn(\bfx)$ can be viewed as a mapping
from the compactified $R^3 \sim S^3 \to S^2$. 
Such mappings fall into nontrivial homotopy 
classes $\pi_3(S^2) \simeq Z$ that can  
be characterized by the Hopf 
invariant \cite{ati}, \cite{kauf}.
For this we introduce the closed two-form 
\be
F =  ( d \bfn \wedge d \bfn , \bfn )
\la{F}
\ee
on the target $S^2$. Since $H_2(S^3) = 0$ 
its preimage 
$F_\star$ on the base $S^3$ is exact, 
\be
F_\star = d A_\star
\la{Fexact}
\ee
and the Hopf invariant $Q_H$ coincides with
the three dimensional Chern-Simons term,
\be
Q_H \ = \ \frac{1}{4\pi^2} 
\int\limits_{R^3} F \wedge A 
\la{hopf}
\ee
The existence of stable solitons in
(\ref{ham1}) with
a nontrivial Hopf invariant
is then strongly suggested by the lower bound estimate 
\be
H  \geq  c \cdot | Q_H |^{\frac{3}{4}}
\la{bound}
\ee
where $c$ is a nonvanishing numerical constant 
\cite{vak}, \cite{kolme}. 

In the following we shall find it useful
to introduce a unit four vector 
$\Phi_\mu(\bfx): R^3 \sim S^3 \to S^3$ and define 
two complex variables
\bastar
Z_1(\bfx) \ & = & \ \Phi_1(\bfx) + i\Phi_2(\bfx) \\
Z_2(\bfx) \ & = & \ \Phi_3(\bfx) + i\Phi_4(\bfx)
\eastar
so that
\[
|Z_1|^2 \ + \ |Z_2|^2 \ = \ 1
\]
and describe the unit three vector $\bfn(\bfx)$ by 
\be
n^k \ = \ Z^{\dagger} \sigma^k Z
\la{ndef}
\ee
with $\sigma^k$ the Pauli matrices.  The two-form 
$F_{ij}$ in (\ref{F}) becomes 
\be
F_{ij} \ = \ i 
(\partial_i Z^\dagger \partial_j Z
-  \partial_j Z^\dagger \partial_i Z)
\la{Fdef}
\ee
and modulo U(1) gauge transformations we have
\be
A_i = \frac{i}{2}( Z^\dagger \partial_i Z
- \partial_i Z^\dagger  Z )
\la{Adef}
\ee
Under a local U(1) rotation
$Z \to e^{i \gamma} Z$ both $F_{ij}$ and
$\bfn$ remain invariant while $A_i$ suffers a gauge
transformation $A_i \to A_i + \partial_i \gamma$.
Hence $Q_H$ is a functional of $\bfn$ 
only, and substituting in (\ref{hopf}) we get
\be
Q_H \ = \ \frac{1}{12\pi^2} \int d^3 x \ 
\epsilon_{\mu\nu\rho\sigma}
\Phi_\mu \ d \Phi_\nu \wedge d \Phi_\rho 
\wedge d \Phi_\sigma
\la{hopf2}
\ee
which is the standard integral 
representation 
of the winding number for maps $S^3 \to S^3$,
invariant under arbitrary local 
variations $ \Phi_\mu \to \Phi_\mu + 
\delta \Phi_\mu$. In particular,
(\ref{hopf}) is an integer valued 
invariant of the 
vector field $\bfn(\bfx)$.

\vskip 1.0cm

\section{Toroidal Solitons}

In the following we shall be interested 
in the simplest possible 
knotlike soliton of (\ref{ham1}),
the torus-shaped unknot configuration \cite{fadde1},
\cite{kolme}. 
In this case we can introduce further 
simplifications using the fact that the 
Hamiltonian (\ref{ham1}) is invariant under both 
spatial SO(3) rotations and internal
SO(3) rotations of the vector $\bfn(\bfx)$. Since 
we expect a torus shaped soliton to exhibit 
rotation invariance around its toroidal symmetry 
axis, this global 
SO(3)$\times$SO(3) invariance will be broken
into a diagonal 
SO(2) $\in$ SO(2)$\times$SO(2)$~$ invariance of 
combined spatial and internal rotations 
around the symmetry axis. 
Instead of minimizing the energy of the original 
Hamiltonian with respect to arbitrary variations 
of $\bfn(\bfx)$, it is then sufficient to minimize 
the energy for the most general toroidal
SO(2) $\in$ SO(2)$\times$SO(2) invariant 
configuration $\bfn(\bfx)$. 

We select the toroidal symmetry axis
so that it coincides with the $z$-axis in 
$R^3$, and align the asymptotic vector field
$\bfn_0$ so that it points along the positive
$z$-axis. In terms of
cylindrical coordinates $(r,z,\psi) \equiv (\bfy , 
\psi)$, the
most general Ansatz which is consistent with 
the toroidal SO(2) symmetry 
then separates the angle $\psi$ 
that describes rotations around the $z$-axis,
\be
\bfn(\bfy , \psi) \ = \
\left( \begin{array}{c} \sin \bigl( \varphi(\bfy) + 
k\psi \bigr) 
\cdot \sin \theta(\bfy) \\
\cos \bigl( \varphi(\bfy) + k\psi \bigr) \cdot
\sin \theta(\bfy) \\
\cos \theta(\bfy) \end{array} \right)
\la{yrite}
\ee
Here $k$ is an integer that counts the number of
times the torus circles around the $z$-axis. 
Since (\ref{yrite}) approaches the asymptotic 
constant vector 
$\bfn_0$ when $|\bfx| \to \infty $ and
$\bfn_0$ points along the 
positive $z$-axis, we conclude that
\be
\theta (\bfy) \ \stackrel{|\bfy| \to 
\infty}{\longrightarrow}  \ 0 \ \ \ \ mod 
\ \ \ 2\pi 
\la{thetalim1}
\ee

We substitute the Ansatz (\ref{yrite})
in (\ref{ham1}) and scale $H$ appropriately
by the dimensionless coupling constant $g$.
In this way we find
for the Hamiltonian
\[
H \ = \ G^2 \cdot E_2 \ + \ E_4 \ = \ 
\int dr dz \ r \cdot {\cal E}(r,z)
\]
\be
=  \int dr dz \ r \Biggl( 
\frac{G^2}{4} ( \theta_r^2 + \theta_z^2 )
\ + \sin^2 \theta \cdot \Bigl[
\frac{G^2}{4} \Bigl\{ \varphi_r^2 
+ \varphi_z^2 + \frac{k^2}{r^2} \Bigr\} \ + \
\frac{1}{16 } \Bigl\{ ( \varphi_r \theta_z 
- \varphi_z \theta_r)^2 + \frac{k^2}{r^2} 
(\theta_r^2 + \theta_z^2) 
 \Bigr\} \Bigl] \Biggr) 
\la{kulmaham}
\ee
Here $G$ is a dimensionfull coupling constant 
inversely proportional to length. 
Since there are no additional 
dimensionfull quantities,
$G$ specifies a length scale which
determines both the size and the shape 
of the solitons. We note 
that $H$ is invariant under $z \to -z$
reflections.
 
We introduce the 
following parametrization of the
four-vector $\Phi(\bfx)$,
\be
\Phi \ = \ \left( \begin{array}{c}
\cos \phi_{12} \sin \vartheta \\
\sin \phi_{12} \sin \vartheta \\ 
\cos \phi_{34} \cos \vartheta \\
\sin \phi_{34} \cos \vartheta  
\end{array} \right)
\la{4vec}
\ee
From (\ref{ndef}), (\ref{yrite}) we then find
\ba
\varphi(\bfy) + k \psi \ & = & \ 
\phi_{34} - \phi_{12}  \non\\
\theta(\bfy) \ & = & \ 2 \vartheta
\la{kulmat1}
\ea
For the gauge field (\ref{Adef})
\be
A \ = \ \cos^2 \vartheta \ d \phi_{34} \ + \ \sin^2 
\vartheta \ d \phi_{12}
\la{Adef2}
\ee
and (\ref{F}), (\ref{Fdef}) yields
\be
F \ = \ dA \ = \ \sin 2\vartheta \ d\vartheta 
\wedge ( d\phi_{34} - d \phi_{12})
\la{F2}
\ee
so that the Hopf invariant (\ref{hopf}) becomes
\be
Q_H \ = \ \frac{1}{\pi^2} \int \sin 2\vartheta \ 
d\vartheta \wedge d\phi_{34} \wedge d\phi_{12}
\la{hopf3}
\ee
Here the expected local $S^1 \times S^2$ structure 
of our toroidal configuration is manifest. 

From (\ref{kulmaham}) we conclude that for a 
finite energy the angle $\theta = 2 \vartheta$ 
must vanish on the $z$-axis, 
\be
\theta(r = 0 , z) \ = \ 0 \ \ \ \ mod \ \ \ 2 \pi
\la{thetalim2}
\ee
Together with (\ref{thetalim1}) this implies that
we can select $\theta = 0$ on the entire 
boundary of the half-plane $(r,z)$. 
In order to have a nontrivial Hopf 
invariant we need
that at some interior points $(r_c,z_c)$ 
of the half plane we have $\theta(r_c,z_c) = \pi
\ \ mod \ \ 2\pi$.
For simplicity we assume that there is only one such 
point, and due to the $z \to -z$ symmetry 
of (\ref{kulmaham}) we can select $z_c = 0$. 
This point determines the 
center of the toroidal configuration, a 
circle of radius $r_c$ on the $z=0$ plane of $R^3$
which is parametrized by the angle $\psi$. 

In (\ref{4vec}) the preimage of $\theta = 0$
corresponds to a circle which is 
parametrized by the angle $\phi_{34}$.
Similarly, the preimage of $\theta = \pi $ 
corresponds to a circle parametrized by the 
angle $\phi_{12}$. Consequently
we can refine (\ref{kulmat1}) by the further
identifications
\ba
\varphi(\bfy) \ & = & \ \ \phi_{34} \non\\
k \psi \ \ & = & \ - \phi_{12}
\la{kulmat2}
\ea
and for the Hopf invariant we obtain 
\be
Q_H \ = \ \frac{k}{2\pi^2} \int \sin \theta \ 
d\theta \wedge d\varphi \wedge d\psi
\ = \ \frac{k}{\pi} \int \sin \theta \ 
d\theta \wedge d\varphi 
\la{hopf4}
\ee
For a nontrivial Hopf invariant
the angle $\varphi(\bfy)$ must then increase 
(decrease) by $2\pi$ (more generally 
by $2\pi n$ with $n$
an integer) when we go once around the
center $r_c$ on the $(r,z)$ half plane. This means
that the two circles parametrized by $\phi_{12}$
and $\phi_{34}$ corresponding to 
the preimages of $\theta = \pi$
and $\theta = 0$ respectively, 
are linked ($k\cdot n$ times).

Finally, if we assume that our toroidal soliton
solution is an analytic function of the
variables $r$ and $z$, we conclude
from (\ref{kulmaham}) that its energy density 
vanishes at the center of the torus
${\cal E}(r_c,0) = 0$, where $\theta$ has a maximum
value $\theta = \pi$.
Since the minimum value $\theta = 0$ occurs
on the boundary of the $(r,z)$ half 
plane, for an analytic 
soliton the energy density must also 
vanish there, and in particular 
along the $z$-axis. Hence ${\cal E}(r,z)$ 
must have a maximum value on a
ring between the center $r_c$ and the boundary of the 
half plane. This means that the energy density of a toroidal
soliton is concentrated 
in a tube-like neighborhood around 
its center.

In the following we find it convenient
to introduce stereographic coordinates
$U(\bfy)$, $V(\bfy)$ on the $(r,z)$ 
Riemann sphere, defined with respect to the
north pole at $\theta = 0$
\be
\varphi =  - \arctan ( \frac{V}{U} ) 
\la{phikulma}
\ee
\be
\theta =  2 \arctan \sqrt{ U^2 + V^2 } 
\la{thetakulma}
\ee
In these variables the 
Hamiltonian (\ref{kulmaham})
becomes
\[
H \ = \ G^2 \cdot E_2 \ + \ E_4 \ = \ \int dr dz r 
\cdot {\cal E}(r,z)
\]
\[
= \ \int drdz \ r \Biggl\{ G^2 \Bigl( 
{ U_{r}^2 + U_z^2 + V_r^2 + V_z^2 
\over ( 1 + U^2 + V^2 )^2 } \ + \ \frac{k^2}{r^2}
{ U^2 + V^2 \over
( 1 + U^2 + V^2 )^2 } \Bigr) 
\]
\be
+ \ { ( U_r V_z - U_z V_r )^2 
\over ( 1 + U^2 + V^2 )^4 }
\ + \ \frac{k^2}{r^2} { 
(U U_r + V V_r )^2 + (U U_z + V V_z )^2
\over (1 + U^2 + V^2)^4 } \Biggr\}
\la{2dham}
\ee
and for the Hopf invariant we get
\be
Q_H \ = \ - \frac{k}{\pi} \int { dU \wedge dV \over
(1 + U^2 + V^2)^2 }
\la{unhopf2}
\ee

\vskip 1.0cm
 
\noindent
\section{Flow Equation}

The Euler-Lagrange equations for (\ref{kulmaham}) 
are highly nonlinear and an analytic solution 
appears to be impossible. It seems to us, that the 
only tools available are numerical.
However, even a numerical integration 
is quite nontrivial, there are several 
complications that need to be resolved.
One concerns the (expected) chaotic
nature of the equations of motion: 
We do not expect a direct Newton's 
iteration to converge towards a stable 
configuration unless we succeed in
constructing an initial configuration
which is very close to the actual solution. 
Rather, we expect Newton's iteration to exhibit 
chaotic behavior. 

Due to lack of a proper initial
configuration for Newton's iteration, we
resort to alternative methods. For this 
we first formulate the problem at an abstract
level, by considering a generic static 
energy functional $E(q)$ with some variables $q_a$. 
We introduce an auxiliary 
variable $\tau$, and extend the (stationary)
Euler-Lagrange equations of $E(q)$ 
to the following parabolic
gradient flow equation
\be
{ d q_a \over d\tau } \ = \ - { \delta E \over
\delta q_a}
\la{flow1}
\ee
Since
\be
{\partial E \over \partial \tau } \ = \ - \biggl( 
{\delta E \over \delta q_a } \biggr)^2
\la{flow2}
\ee
the energy decreases along 
the trajectories of (\ref{flow1}). Furthermore,
by squaring (\ref{flow1})
and integrating from some initial value $\tau = T$ to
$\tau \to \infty$ we get
\be
\int\limits_T^\infty d\tau 
\biggl( { d q_a \over d\tau } \biggr)^2
\ = \  \int\limits_T^\infty d\tau \biggl(
{ \delta E  \over \delta q_a} \biggr)^2
\la{flowint}
\ee
Hence $\tau$-bounded trajectories 
of (\ref{flow1}) flow towards a stable critical
point of $E(q)$. In particular, 
if we start at initial time $\tau = T$ from
an initial configuration $q = q_0$  and follow 
a bounded trajectory of (\ref{flow1}), in the 
$\tau \gg T$ limit we eventually flow to a stable
critical point of $E(q)$ 

In (\ref{2dham}) the fields
$U(\bfy)$, $V(\bfy)$ correspond to the variables $q_a$,
and by denoting $W_1 = U$, $W_2 = V$ the
flow equation becomes 
\be
\frac{ \partial W_a(\bfy)}{\partial \tau} \ = \ 
- G^2 \cdot \frac{ \delta E_2 }{\delta W_a(\bfy)} 
\ - \  \frac{ \delta E_4 }{\delta W_a(\bfy)} 
\la{flow3}
\ee
If we introduce the scaling (\ref{scalex})
we find for the scaled Hamiltonian the flow equation
\be
\frac{ \partial W_a(\bfy)}{\partial 
(\frac{1}{\rho} \tau )} \ = \ 
- (\rho^2 G) \cdot \frac{ 
\delta E_2 }{\delta W_a(\bfy)} 
\ - \ 
\frac{ \delta E_4 }{\delta W_a(\bfy)} 
\la{scaleflow1}
\ee
Thus a flow towards a toroidal soliton 
with coupling constant $G$ coincides
with the flow towards a toroidal soliton 
with coupling constant $ \rho^2 G$,
provided we rescale the flow 
variable $\tau$ into 
$\frac{1}{\rho} \tau$. 
This means that the soliton is essentially unique;
It is sufficient to 
consider the flow towards a soliton 
with a definite value for $G$, since solitons 
with other values of $G$  
are obtained from this configuration
by a simple scale transformation.
Notice however, that in practice we 
perform a numerical integration
of (\ref{flow3}) on a finite lattice, 
and there are finite size corrections to the  
simple scaling (\ref{scaleflow1}).

An additional problem in a numerical approach
is the selection of an optimal size for the lattice.
If the lattice is too small in comparison to the
scale that describes the soliton, the soliton
may not fit into it. On the other hand, 
if the lattice is too large in comparison to the
soliton, we may either
miss the soliton entirely or use  
an unnecessarily large amount of computer time in 
constructing it.

The coupling constant 
$G$ in (\ref{2dham}) is the sole 
dimensionfull quantity that appears in
our equations. Consequently it determines how 
both the location $r_c$ of the center and the
thickness of the soliton scale. 
Besides $G$, there are also dimensionless
numerical factors that affect the overall
size of the soliton. Since these numerical 
factors can only be obtained by actually 
solving the equations of motion,
there are no {\it a priori} methods 
in selecting an optimal size of a 
lattice. 

In order to approach these problems, 
we have developed a simple renormalization 
procedure that allows us to select the 
initial configuration so that the location 
$r_c$ of its center coincides with the location
of the center for the actual soliton. 
For this we first observe
that the soliton obeys the virial theorem 
(\ref{virial}), in our present variables
\[ 
G^2 E_2 \ = \ E_4
\]
By demanding that this virial 
theorem is also obeyed during the 
flow (\ref{flow3}), we promote the 
coupling constant $G$ into a $\tau$-dependent 
variable $G \to G(\tau)$,
\[
G(\tau) \ = \ \sqrt{ { E_4(\tau) \over E_2(\tau) } }
\]
When we approach a soliton as 
$\tau \to \infty$, 
the variable $G(\tau)$ must then 
flow towards an 
asymptotic value $G^*$ which is the  
value of the coupling for the soliton, 
\be
G(\tau) \stackrel{\tau \longrightarrow 
\infty}{\rightarrow}  G^*
\la{renorm}
\ee
This renormalization procedure fixes
the location $r_c$ of the center for the 
soliton, and allows us to choose 
the size of our lattice appropriately.  

The thickness of the final 
soliton is also determined by the asymptotic value 
$G^*$ of the coupling constant.
However, since additional
dimensionless numerical
factors are also 
involved, the renormalization (\ref{renorm}) 
does not help us in selecting the 
thickness of the initial condition so that
numerical convergence is secured. 
This poses a problem for which we 
at the moment lack a firm solution,
besides experimenting with various different 
types of initial configurations.    

Finally, on a finite size lattice we also need to 
determine boundary conditions at the boundary
of the lattice. On the $(r,z)$ half plane
the proper boundary condition for 
$U(r,z)$ and $V(r,z)$ is that both vanish 
on the boundary of the half plane.  
This corresponds to the
compactification of the half plane into 
a Riemann sphere. However, we expect that on a finite 
lattice a simulation with such trivial
boundary conditions 
leads to a flow towards the trivial configuration 
$U(r,z) \equiv V(r,z) \equiv 0$. 
In order to impose the boundary conditions
properly in our numerical simulation, 
we have adapted an iterative process 
where we first specify the boundary conditions
using the initial configuration $U_0(\bfy)$, $V_0(\bfy)$. 
At later values of $\tau$ we then update 
these boundary conditions successively,
by interpolating the iterated configurations 
from the interior of the lattice to its boundary. 
In this manner we expect that we eventually 
obtain boundary conditions that correspond to 
those of an actual toroidal soliton.
An alternative method that we have used to 
determine the boundary 
conditions, is to start the iteration of (\ref{flow3})
from a sufficiently large initial lattice with 
boundary conditions determined by the initial configuration.
By successively shrinking the size of the lattice and 
determining boundary conditions in the shrinked 
lattice using the appropriate restriction of the
pertinent iterated configuration from the larger lattice,
we expect to converge towards boundary 
conditions that coincide with those of an 
actual soliton.

\vskip 1.0cm

\section{Numerical Simulation}

We shall now present an example of a numerical
construction of an unknot soliton. For simplicity
we restrict to a configuration with Hopf invariant
$Q_H = 1$ ({\it i.e.} $k = n = 1$), 
the other cases being treated similarly. 

For a numerical integration of 
(\ref{flow3}), we need an initial
configuration with $Q_H = 1$ that 
properly approximates the final soliton. For this
we introduce the standard complex bilinear 
transformation
\be
\xi \ = \ { \sigma -1  \over \sigma + 1 }
\la{abc}
\ee
and view it as a map between the unit 
disk $| \xi | \leq 1 $ and the $Re(\sigma) 
\geq 0$ half plane. By employing polar coordinates
on the disk, we observe that 
the loci on the $w$-plane are circles, that at least 
in a qualitative sense coincide with our {\it a priori}
expectations for the angular (\ref{phikulma})
and radial (\ref{thetakulma}) behavior of
a soliton (in coordinates where 
the center of a soliton coincides
with $\theta = 0$).

Our soliton  maps  
our compactified $(r,z)$ half plane to a 
Riemann sphere. The  variables 
$U(r,z)$ and $V(r,z)$ are stereographic coordinates 
with respect to the north pole $\theta = 0$, 
which topologically coincides with the
entire boundary of the $(r,z)$ half plane. 
As a consequence $U$ and $V$ determine a mapping 
from the $(r,z)$ half plane to a disk, with
the center of the disk at $\theta = 0$ 
(the entire boundary of the $(r,z)$ half plane),
while the rim of the disk is an (infinitesimally) 
small circle around the center of the 
soliton at $r = r_c$ where we have $\theta = \pi$.
This suggests that we start our construction
of an initial configuration by first setting
\be
W(r,z) \ = \ U(r,z) \ + \ i V(r,z) \ = \ { r + i z + \rho 
\over r + iz - r_c }
\la{init1}
\ee
so that
\ba
U(r,z) \ & = & \ 1 \ + \ (r_c 
+ \rho) \cdot { r - r_c \over 
(r - r_c)^2 + z^2 }
\non\\
V(r,z) \ & = & \ - (r_c + \rho) \cdot { z \over 
(r - r_c)^2 + z^2 }
\la{init2}
\ea
and the loci are circles on the $U$, $V$ plane with
\[
( U - 1 )^2 \ + \ ( V 
+ \frac{ r_c + \rho }{ 2 z } )^2 \ = \
\frac{ (r_c + \rho)^2}{  4 z^2 }
\]
\[
(U - \half - \half \frac{ r + \rho }{r - r_c} )^2
\ + \ V^2 \ = \ \frac{1}{4} \frac{ (r_c + \rho)^2}
{(r - r_c)^2 }
\]
Here $\rho > 0$ is a parameter, according to
(\ref{abc}) the canonical value is $\rho = r_c$ but
for the moment we leave it unspecified. 

For (\ref{init1}) we 
have $\theta(r_c,0) = \pi$ in (\ref{thetakulma}). 
But in addition we need $\theta = 0$ on the entire 
boundary of the $(r,z)$ half plane. For this we 
improve (\ref{init1}) into  
\ba
U(r,z) \ & = & \ F_U(r,z) \cdot \Biggl(
1 + (r_c + \rho) \cdot { r - r_c \over 
(r - r_c)^2 + z^2 } \Biggr) 
\non\\
V(r,z) \ & = & \ - F_V(r,z) \cdot \Biggl(
(r_c + \rho) \cdot { z \over 
(r - r_c)^2 + z^2 } \Biggr)
\la{alku1}
\ea
where $F_U$, $F_V$ are appropriate
positive valued functions, that vanish on the boundary of
the $(r,z)$ half-plane; For a proper 
initial configuration both $F_U$ and $F_V$ 
vanish sufficiently rapidly when 
we move away from the center at $r = r_c$,
so that the initial configuration (in addition
of the final soliton) can be fitted inside the 
lattice that we use in our simulation. 
We also 
need to specify these functions so that the 
Hopf invariant of the initial configuration 
coincides with the Hopf invariant of the desired
soliton, in the present case $Q_H = 1$.
 
\vskip 0.5cm
We have performed extensive numerical 
simulations for a soliton centered at $r_c = 6$,
using several different initial profiles $F_U$
and $F_V$, lattice structures and various alternative 
methods for ensuring independence from 
boundary behavior. Our simulations are consistently 
converging towards a definite toruslike configuration.
As an example we now describe results from a 
particular simulation, based on the initial configuration 
(\ref{alku1}) with
\be
F_U(r,z) \ = \ F_V(r,z) \ = \ b \cdot
{ \exp\bigl( - d_1 (r - r_c)^2
- d_2 z^2 \bigr)  \over c_1 + 
(r-r_c)^2 + c_2 z^2 } \cdot
( 1 - \tanh[
\sqrt{ e_1 ( r - r_c)^2 + e_2 z^2 } ] )
\la{inimod}
\ee
where $b , ... , e_2$ are parameters, 
fixed by minimizing the total energy 
(\ref{2dham}) with respect to these 
parameters. 
This minimization is subject to the supplementary
condition, that the overall size and shape 
of the initial configuration should resemble 
as much as possible the fixed point configurations 
that we have found in our 
earlier numerical simulations. In particular, we use 
the angular ($\varphi$) profiles of our earlier
simulations to select an optimal value for the
parameter $\rho$ in (\ref{init1}).

In a simulation of the flow equation (\ref{flow3})
we determine the length of
each time step $\Delta \tau \_n = \tau_{n} - \tau_{n-1}$ 
adaptively, by demanding that 
the relative variation of total energy remains 
bounded,
\be
| { \Delta E(\tau_n) \over E(\tau_n) } | \ \leq \ 10^{-4}
\la{relativ}
\ee

We employ version 5.4 of the 
PDE2D finite element algorithm \cite{sewell}
using a sparse direct method with 
a 4$^{th}$ order Hermitean polynomial finite 
element basis. We have chosen a finite 
element approach, since it computes a 
continuous piecewise polynomial approximation 
to the solution. In a problem of topological 
nature this should be a definite 
advantage {\it e.g.} over a 
finite difference approach, where 
the solution is approximated only at discrete 
lattice nodes.
In our example we have started from an initial 
square lattice with $0.001 \leq r \leq 14.0$ and 
$-6.0 \leq z \leq 6$, divided into a finite element 
mesh with 15.000 
triangular elements. We have selected the 
triangulation so that it is more dense 
near $r \approx 0$ and $z \approx 0$, where 
we expect to have strongest dependence on the initial 
boundary conditions. 

On the initial lattice 
we iterate the flow equation (\ref{flow3}) until 
we reach a fixed point configuration 
described in figures 1a, 2a, 3a. 
We then introduce a sublattice 
with $ 0.1 \leq r \leq 13.5$ and $ -5.5 \leq z \leq 5.5$, 
divided into a finite element mesh with 18.000 evenly 
distributed triangular elements. On this sublattice 
we determine both the initial and boundary conditions 
by restricting the previously obtained fixed point 
configuration to the sublattice. We then continue 
the iteration of the flow equation (\ref{flow3}) 
until we again reach a fixed point, described in
Figures 1b, 2b, 3b and 4-7.

In this two-phase simulation each of the 
phases takes about 200 hours of CPU time, on 
a Digital Alpha Server 8400 equipped with EV56/440 MHz 
processors and 4GB of CPU. In the second phase with a
larger number of triangles, our computation consumes 
about 725MB of CPU.

Our simulations reveal very strong convergence 
both for the coupling constant $G(\tau)$ and for
the total energy $E(\tau)$ towards fixed point 
values; see figures 1-2. The length of the time 
step also increases exponentially when we approach 
the fixed point; see figure 3. Furthermore, we 
find very strong pointwise convergence for both 
of the angular functions $\theta(r,z)$ and 
$\varphi(r,z)$, and also for the density 
$Q_H(r,z)$ of the Hopf invariant; 
see figures 4-6. The integrated Hopf invariant $Q_H$ 
is very stable under our entire simulation, 
essentially $Q_H \approx 1$ and for the final 
configuration we have $Q_H \ = \ 0.99997 ...$, when 
integrated over the final lattice.

Finally, for the energy 
density  ${\cal E}(r,z)$ in (\ref{2dham}) 
we have also found very 
strong pointwise convergence 
except in the vicinity of 
the origin $r \approx  z \approx 0$. In this
region we experience difficulties in fully
eliminating the residual dependence of
${\cal E}(r,z)$ from our initial boundary 
condition along the $z$-axis. 
This slowness in pointwise convergence 
for very small $r$ and $z$ in ${\cal E}(r,z)$
can be related to our 
method: We use a finite element 
approach, where in each triangle of the mesh 
we approximate the solution by a 
4$^{th}$ order Hermitean polynomial 
in $r$ and $z$. Since ${\cal E}(r,z)$
involves $r^{-2}$ terms, it is then conceivable 
that if the derivatives of $U(r,z)$ and 
$V(r,z)$ in (\ref{2dham}) do not vanish 
sufficiently rapidly for small $r$,
the pointwise 
convergence of ${\cal E}(r,z)$ may become 
slow in this region.

However, if we include the additional 
measure factor $r$ 
in (\ref{2dham}), it tames the $r^{-2}$ behavior
to the extent that the integrand of the total
energy does exhibit very strong pointwise convergence. 
This is also the reason, why the difficulties we experience 
in eliminating the boundary dependence from 
${\cal E}(r,z)$ as $r,z \to 0$ do not interfere 
with the overall convergence of the energy for
our solution.

In figure 7 we present the energy density 
${\cal E}(r,z)$ of our 
final configuration for $r \geq 0.3$, which
is the region where we have confidence in
the pointwise convergence of ${\cal E}(r,z)$
in the present example. The result is consistent 
with our qualitative picture: The energy 
density has a definite tubelike shape, essentially  
vanishing inside a tubular region 
around the center at $r = r_c$ and then attaining 
a maximum value in a surrounding 
collar-like region. At $z=0$ the maximum value of 
the energy density is obtained for a relatively  
small value of $r=r_{max}$, and 
by combining results from a number of 
different simulations we have arrived at
an estimate that $r_{max} \leq 0.9$. 
For smaller values of $r$ the energy 
density ${\cal E}(r,z=0)$ then appears to
decrease very rapidly, but due to 
the ensuing difficulties with small
$r$ convergence we can not verify that on
the $z$-axis we actually have 
${\cal E}(r=0,z) \approx 0$. For this one needs to
perform a high precision three dimensional 
simulation, such a simulation is now in progress
and we plan to report on it in a
future publication.

\vskip 1.0cm

\section{Conclusions}

In conclusion, we have analyzed in detail 
the torus shaped unknot soliton using a high 
precision simulation. We have found very 
strong convergence towards
a critical point of the energy, suggesting that
we have indeed found a very good aproximation to 
an actual soliton. Our results indicate, that
a combination of geometrical and topological tools
in quantum field theory with high performance
computing is finally making a first principles
investigation of knotlike solitons realistic.
In particular, our results indicate that
Kelvin's conjecture on the existence of stable 
torus knots can be realized in the model 
that we have studied.

\vskip 1.5cm

We thank J. Hietarinta, A.P. Niemi,
J. Pitk\"aranta, G. Sewell and S. Virtanen 
for discussions, and in particular 
K. Palo for discussions and helping us 
with scripts. We are also grateful for
the Center for Scientific Computing in Espoo,
Finland for providing us with an access to 
their Digital Alpha Server 8400 computer.

\vfill\eject

\vfill\eject
\begin{flushleft}
{\bf Figure Caption}
\end{flushleft}
\vskip 1.0cm

{\bf Figure 1a:} Flow of coupling constant $G(\tau)$
towards its fixed point value $G^*$, for a soliton
with $r_c = 6$.

\vskip 0.4cm
{\bf Figure 1b:} Flow of coupling constant $G(\tau)$
towards its fixed point value $G^*$, starting from the
fixed point configuration described in Fig. 1a on a
truncated sublattice, with boundary conditions also
determined by the fixed point configuration in Fig. 1a.
The slight difference in the 
fixed point values $G^*$ after the first and second
phase is partly due to a smaller and 
relatively more accurate lattice 
employed in the second run, and partly due to an 
improvement in the boundary conditions. Due to 
improved convergence there
is also a ${\cal O}(10^3)$ increase in the final
time step scales.

\vskip 0.4cm
{\bf Figure 2a:} Flow of total energy as a function
of $\tau$ towards a fixed point. Notice that the energy 
decreases as a function of $\tau$, as it should for a 
bounded trajectory.

\vskip 0.4cm
{\bf Figure 2b:} Flow of total energy as a function
of $\tau$ towards a fixed point, starting from the
fixed point configuration described in Fig. 2a on
a truncated sublattice, with boundary conditions
determined by the fixed point configuration in Fig. 2a.
Notice that there is a  ${\cal O}(10^3)$ improvement
in the time step scales.

\vskip 0.4cm
{\bf Figure 3a:} Length of time step as a function of
the number of iterations. The length of a time 
step is determined by the bound (\ref{relativ}). 
The length should increase without a limit when a 
fixed point is approached. 

\vskip 0.4cm
{\bf Figure 3b:} Length of time step for the second phase,
starting from the configuration described in Figure 3b.
The relative increase by a factor of ${\cal O}(10^3)$
from Fig. 3a can be attributed to improvements 
in boundary behavior and lattice accuracy. 

\vskip 0.4cm
{\bf Figure 4:} The angle $\theta(r,z)$ for a fixed
point configuration at the end of the second phase.
Plotted on a sublattice with $0.3 \leq r \leq 12$ and
$-4.5 \leq z \leq 4.5$. At the center $r_c = 6$
we have $\theta = \pi$, and outside of the soliton 
$\theta \approx 0$.

\vskip 0.4cm
{\bf Figure 5:} The angle $\varphi(r,z)$ for a fixed
point configuration at the end of the second phase.
Plotted on a sublattice with $0.3 \leq r \leq 10$ and
$-4.5 \leq z \leq 4.5$. When we go once around the
center at $r_c = 6$, $\varphi$ jumps by $2\pi$.
Notice the appearance of a tubular structure in
the middle, corresponding to the interior of the
soliton.

\vskip 0.4cm
{\bf Figure 6:} The density $Q_H(r,z)$ of the Hopf
invariant, for the fixed point configuration
at the end of the second phase. The 
tubular structure of a soliton is clearly visible.
The integrated value of the Hopf invariant is very
stable under our entire iteration, and for the
final configuration we have $Q_H = 0.999997 ...$

\vskip 0.4cm
{\bf Figure 7:} The energy density ${\cal E}(r,z)$ 
for the fixed point configuration at the end of
the second phase, plotted on a sublattice with 
$0.3 \leq r \leq 10$ and $-4.5 \leq z \leq 4.5$.
The tubular structure is clearly visible, but near
$r \approx z \approx 0$ we still have residual
dependence on the initial boundary on the $z$-axis.
This is understandable, since even a small error
in the boundary condition can eventually 
give rise to a large local tension.

\end{document}